\documentclass[floatfix,twocolumn,aps,showpacs,showkeys,preprintnumbers,nofootinbib]{revtex4}
\usepackage{amssymb}
\usepackage{amsmath}
\usepackage{graphics}
\usepackage{epsfig}
\usepackage[dvips]{color}
\usepackage{bm}
\usepackage{color} 
\newcommand{\be}{\begin{equation}}

\newcommand{\ee}{\end{equation}}
\newcommand{\bea}{\begin{eqnarray}}
\newcommand{\eea}{\end{eqnarray}}
\newcommand{\beas}{\begin{eqnarray*}}
\newcommand{\eeas}{\end{eqnarray*}}

\newcommand{\nn}{\nonumber\\}
\newcommand{\slsh}[1]{{\not \! #1}}

\begin{document}
\setlength{\parskip}{\baselineskip} 
\thispagestyle{empty}
\pagenumbering{arabic} 
\setcounter{page}{1}
\mbox{ }
\preprint{UCT-TP-304/15}
\title{Quark deconfinement and gluon condensate in a weak magnetic field\\
 from QCD sum rules}
\author{Alejandro Ayala} \affiliation{Instituto de Ciencias
  Nucleares, Universidad Nacional Aut\'onoma de M\'exico, Apartado
  Postal 70-543, M\'exico Distrito Federal 04510,
  Mexico}
  \affiliation{Centre for Theoretical and Mathematical Physics, and Department of Physics,
    University of Cape Town, Rondebosch 7700, South Africa}
\author{C. A. Dominguez}\affiliation{Centre for Theoretical and Mathematical Physics, and Department of Physics,
  University of Cape Town, Rondebosch 7700, South Africa}
  \author{L. A. Hernandez}\affiliation{Centre for Theoretical and Mathematical Physics, and Department of Physics,
      University of Cape Town, Rondebosch 7700, South Africa}
\author{M. Loewe}\affiliation{Instituto de F\1sica, Pontificia Universidad Cat\'olica de Chile,
      Casilla 306, Santiago 22, Chile}
   \affiliation{Centro Cient\1fico-Tecnol\'ogico de Valpara\1so, Casilla 110-V, Valpara\1so, Chile}
   \affiliation{Centre for Theoretical and Mathematical Physics, and Department of Physics,
    University of Cape Town, Rondebosch 7700, South Africa}
 \author{Juan Cristobal Rojas}\affiliation{Departamento de F\1sica, Universidad Cat\'olica del Norte, Casilla 1280, Antofagasta, Chile}

\author{ Cristian Villavicencio}\affiliation{Departamento de Ciencias B\'asicas, Universidad del B\'io B\'io, Casilla 447, Chill\'an, Chile}

\begin{abstract}

We study QCD finite energy sum rules (FESR) for the axial-vector current correlator in the presence of a magnetic field, in the weak field limit and at zero temperature. We find that the perturbative QCD as well as the hadronic contribution to the sum rules get explicit magnetic field-dependent corrections and that these in turn induce a magnetic field dependence on the deconfinement phenomenological parameter $s_0$ and on the gluon condensate. The leading corrections turn out to be quadratic in the field strength. We find from the dimension $d=2$ first FESR that the magnetic field dependence of $s_0$ is proportional to the absolute value of the light-quark condensate. Hence, it increases with increasing field strength. This implies that the parameters describing chiral symmetry restoration and deconfinement behave similarly as functions of the magnetic filed. Thus, at zero temperature the magnetic field is a catalysing agent of both chiral symmetry breaking and confinement.  From the dimension $d=4$ second FESR we obtain the behavior of the gluon condensate in the presence of the external magnetic field. This condensate also increases with increasing field strength. 

\end{abstract}

\pacs{25.75.Nq, 11.30.Rd, 11.55.Hx}

\keywords{Finite energy sum rules, quark condensate, gluon condensate, pion decay constant, magnetic fields}

\maketitle

\section{Introduction}\label{I}

Lately, the properties  of strongly interacting matter in the presence of external magnetic fields has become a very active research field. One of the driving motivations behind this interest is the possibility to  study experimentally such properties in peripheral collisions of heavy nuclei at high energy. In addition, recent lattice QCD (LQCD) results show that the critical temperature for deconfinement/chiral symmetry restoration decreases with increasing field strength~\cite{Fodor}. This behavior is dubbed {\it inverse magnetic catalysis}, and it reveals an unexpected, non-trivial phenomenon: in a thermal environment, near the transition temperature, the presence of a  magnetic field acting on strongly interacting matter hinders  the formation of a quark-anti-quark condensate. LQCD calculations~\cite{Bali} show that the quark condensate does increase with increasing magnetic field at low temperatures. This behavior corresponds to {\it magnetic catalysis}. However, as the temperature increases approaching the crossover region $T\simeq 150$ MeV,  the quark condensate reaches a maximum  value  smaller than  at $T=0$  (for the same field strength).  Subsequently, the condensate decreases as a function of the field strength. Finally, for temperatures above the cross-over values the condensate  decreases monotonically as a function of the magnetic field. Some of the possible scenarios  aiming to understand this behavior include (i) invoking a fermion paramagnetic contribution to the pressure with a sufficiently large magnetization~\cite{Noronha}, (ii) the competition between the valence and sea contributions at the phase transition~\cite{Bruckmann:2013oba} produced by a back reaction of the Polyakov loop, which  depends on the magnetic field~\cite{Ferreira}, (iii) magnetic inhibition due to neutral meson fluctuations in a strong magnetic field~\cite{Fukushima}, (iv) accounting for non-perturbative effects by means of Schwinger-Dyson  Equations and renormalization group analyses~\cite{Mueller,Andersen2,Braun}, (v) a decreasing magnetic field and temperature dependent coupling with~\cite{amlz, Ayala2, Ayala3} and without~\cite{Farias, Ferreira1} plasma screening effects, (vi) the proper account of the gluon contribution during the phase transition~\cite{Bruckmann}, and (vii)  quark anti-screening due to the anomalous magnetic moment of quarks in strong~\cite{Ferrer} and weak~\cite{Fayazbakhsh} fields. On the other hand, this behavior is not obtained in mean field approaches  describing the thermal environment~\cite{Andersen, Fraga, Loewe, Agasian, Mizher, Fraga2}, nor when calculations beyond mean field do not include magnetic effects on the coupling constants~\cite{ahmrv}. For recent reviews see Refs.~\cite{Andersenreview, Miransky}.

Given the dual nature of the QCD phase transition, a pertinent question is to what extent inverse magnetic catalysis is due to the mechanisms of either  chiral symmetry restoration and/or of deconfinement. One way to address this question is to find a relation between deconfinement and chiral symmetry restoration parameters as a function of the magnetic field. Since the transition happens for temperatures in the realm of non-perturbative phenomena, the relation searched for needs to carry non-perturbative information.
An extensively used tool in the context of effective models at finite temperature and zero~\cite{Scoccola} and finite~\cite{Mizher} magnetic field is the Polyakov loop~\cite{Larryoriginal}. When coupled to quark degrees of freedom this loop sheds light on how chiral symmetry and deconfinement behave during the QCD transition as a function of the field intensity. Another non-perturbative tool that does not rely on effective models is that of QCD Finite Energy Sum Rules (FESR). This  approach has been successfully applied both at zero~\cite{QCDSR_review} and at finite temperature~\cite{FESRT_1} to understand hadronic properties. Of particular mention are (i) the prediction, in this framework, of the survival of charmonium and bottonium above the critical temperature~\cite{FESRT_2}, (ii) the temperature behavior of the hadronic width of charmonium and bottonium~\cite{FESRT_2} in qualitative agreement with bottonium results from LQCD~\cite{Aarts}, and (iii) the description of the di-muon spectrum in heavy-ion collisions~\cite{di-muon}, in excellent agreement with data~\cite{Na60} in the region of the rho-meson peak. A key parameter that emerges from this analysis signalling quark-gluon deconfinement is the squared energy threshold, $s_0$, above which the hadronic spectral function is well approximated by perturbative QCD (pQCD). An interesting relation between $s_0$ and the quark condensate $\langle\bar{q}q\rangle$, whereby the former is proportional to the latter has been found, in the absence of a magnetic field and at finite temperature, in \cite{CAD1}, and at finite temperature and density in \cite{abdglr}.

\begin{figure}[t!]
{\centering {
\includegraphics[scale=0.5]{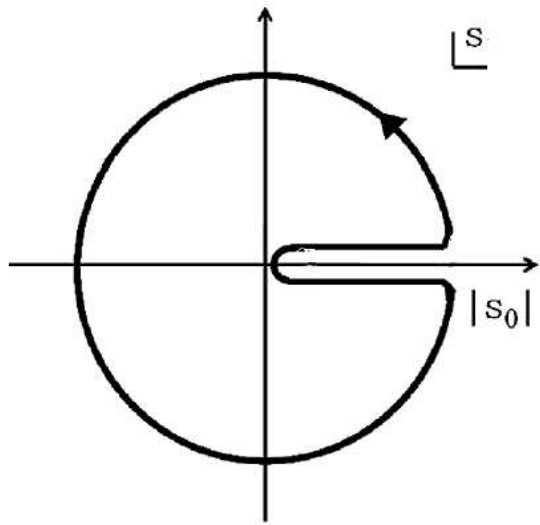}
}}
\caption{Cauchy integration contour in the complex squared energy $s$-plane used to obtain QCD FESR. The radius of the circle, $s_0$, is the threshold for pQCD.}
\label{fig1}
\end{figure}

In this paper we use FESR in the axial-vector channel, and in the presence of an external magnetic field,  to explore the relation between (i) the deconfinement and chiral symmetry restoration parameters, $s_0$ and $\langle\bar{q}q\rangle$,  and (ii) obtain the behavior of the gluon condensate as a function of the magnetic field intensity at zero temperature. Other formulations of the QCD sum rules, also at $T=0$, have been used to explore the dependence of heavy-quark meson-masses and mixing as a function of the magnetic field strength~\cite{others}. The paper is organized as follows: In Sec.~\ref{II} we set up the FESR. We show that since the pQCD contribution to the current correlator receives corrections that can be expressed as powers of the magnetic field strength divided by powers of the  squared energy $s$, there appear additional terms that contribute to higher order FESR. In Sec.~\ref{III} we explicitly compute the  pQCD corrections to the current correlator and solve the first two FESR to find the dependence of $s_0$ and the gluon condensate on the magnetic field strength. We show that for magnetic field strengths $eB$ smaller than $s_0$, the former follows the magnetic field dependence of the quark condensate, which we parametrize from LQCD results~\cite{Bali}. We also show that the magnetic field dependence of the gluon condensate receives non-trivial corrections from the pQCD sector and that overall it is a monotonically increasing function of the field strength. We summarize and conclude in Sec.~\ref{concl}, leaving for the appendices the explicit computation of the imaginary part of the hadronic contribution and the vanishing of the first order correction in the field strength of the pQCD contribution to the axial-vector current correlator.

\section{Finite Energy QCD Sum Rules In the Presence of a Magnetic Field}\label{II}

The charged axial-vector current correlator in the absence of a magnetic field and at $T=0$ can be written as

 \begin{eqnarray}
 \Pi_{\mu\nu} (q^{2})   &=& i \, \int\; d^{4}  x \; e^{i q x} \; 
 <0|T( A_{\mu}(x) \;, \; A_{\nu}^{\dagger}(0))|0> \nonumber \\ [.3cm]
 &=& (-g_{\mu\nu} q^2 + q_\mu q_\nu)\, \Pi_A(q^2) + q_\mu q_\nu\, \Pi_0(q^2),\nn
 \label{axialcorrelator}
 \end{eqnarray}
 where $A_\mu(x) = : \bar{d}(x) \gamma_\mu \, \gamma_5 u(x):$ is the (charged) axial-vector current,  $q_\mu$ is the four-momentum carried by the current, with $s\equiv q^2 >0$ the squared energy. The functions $\Pi_{A,0}(q^2)$ are free of kinematical singularities, an important property needed in writing dispersion relations and sum rules. Concentrating on e.g. $\Pi_0(q^2)$ and invoking the Operator Product Expansion (OPE) of current correlators at short distances beyond perturbation theory, one of the two pillars of the QCD sum rule method, one has
 \begin{equation}
 \Pi_0(Q^2)|_{\mbox{\scriptsize{QCD}}} = C_0 \, \hat{I} + \sum_{N=1} \frac{C_{2N} (Q^2,\mu^2)}{Q^{2N}} \langle \hat{O}_{2N} (\mu^2) \rangle \;, \label{OPE}
 \end{equation}
 where $Q^2 \equiv - q^2$, $\langle \hat{O}_{2N} (\mu^2) \rangle \equiv \langle0| \hat{O}_{2N} (\mu^2)|0 \rangle$, and $\mu^2$ is a renormalization scale. The Wilson coefficients $C_N$ depend on the Lorentz indexes and quantum numbers of the currents, and on the local gauge invariant operators $\hat{O}_{2N}$ built from the quark and gluon fields in the QCD Lagrangian. These operators are ordered by increasing dimensionality and the Wilson coefficients are calculable in pQCD. The unit operator above has dimension $d=0$ and $C_0 \hat{I}$ stands for the purely perturbative contribution normalized according to 
\begin{equation}
C_0\, \hat{I} = \frac{1}{4 \pi} \ln \left( \frac{-s}{\mu^2}\right)\left[ 1 + {\cal{O}} (\alpha_s(s)) \right] \;, 
\end{equation}
Since there are no dimension $d=2$ operators built from the QCD fields, it is generally assumed that the OPE starts at dimension $d=4$. This is fully confirmed by determinations of condensates from experimental data~\cite{d2}. The dimension $d=4$ in the chiral limit is proportional to the renormalization group invariant gluon condensate
\begin{equation}
C_4 \langle \hat{O}_{4}  \rangle = \frac{\pi}{3} \langle\alpha_s \, G^2 \rangle
\end{equation}
The second pillar of the QCD sum rule method is to consider an integration contour in the complex square energy plane, as in Fig. 1, and invoke Cauchy's theorem assuming that QCD can be used on the circle of radius $|s_0|$, provided $|s_0|$ is large enough (quark-hadron duality). On the real axis there is a discontinuity associated with the hadronic states entering the spectral function. Since there are no further singularities this leads to the FESR
\bea
   -\frac{1}{2\pi i}\oint_{C(|s_0|)}\!\!\!\!\!\!\!\!\!\!ds \,s^{N-1}\Pi_0^{\mbox{\tiny{QCD}}}(s)=\frac{1}{\pi}\int_0^{s_0}
   \!\!\!\!\!ds\, s^{N-1}   
   {\mbox{Im}}\,\Pi_0^{\mbox{\tiny{HAD}}} (s),
\label{Cauchy}
\eea
with $N\geq 1$, and $\Pi_0^{\mbox{\tiny{QCD}}}(s)$ given by the OPE, Eq.~(\ref{OPE}).
It will be shown later that in the presence of a magnetic field, and in the weak field limit $eB< s_0$, the Wilson coefficients acquire themselves a B-field dependence. In this work we shall compute the corrections to the FESR due to a weak magnetic field, which can be expressed as a series in powers of $eB$.
Since the magnetic field carries dimension of energy squared, on dimensional grounds one finds the replacements
\bea
   C_{0}\ln\left(\frac{-s}{\mu^2}\right) &\rightarrow& C_0\ln\left(\frac{-s}{\mu^2}\right)  + \sum_{n=1}C_{0}^{(n)}
   \frac{(eB)^n}{s^n}\nn
   C_{2N}&\rightarrow&\sum_{m=0}C_{2N}^{(m)}\frac{(eB)^m}{s^m}
\label{pQCD}
\eea
where $C_{2N}^{(m)}$ are dimensionless quantities that can be computed in pQCD at a given order in $eB$.
Substituting Eqs.~(\ref{pQCD}) and~(\ref{OPE}) into Eq.~(\ref{Cauchy}), one obtains
\bea
   -\sum_{m=0}^{N-1}(-1)^{N-m}C_{2(N-m)}^{(m)}(eB)^m\langle O_{2(N-m)}\rangle&=&\nn
   \frac{1}{\pi}\int_0^{s_0}
   \!\!\!\!\!ds\, s^{N-1}   
   {\mbox{Im}} \,\Pi_0^{\mbox{\tiny{HAD}}} (s)
   -
   \frac{C_0}{N}s_0^N + C_0^{(N)}(eB)^N.&&
\label{SR1}
\eea
Note that in general the presence of the magnetic field mixes operators of different dimension in the FESR. For instance, the first two sum rules $(N=1,2)$ become
\begin{equation}
   0 =\frac{1}{\pi}\int_0^{s_0}
   \!\!\!\!\!ds  \; 
   {\mbox{Im}}\Pi_0^{\mbox{\tiny{HAD}}} (s)   -
   C_0 s_0 + C_0^{(1)}(eB)\;, 
   \label{one}
\end{equation}   
\begin{eqnarray}   
   -C_4^{(0)}\langle O_{4}\rangle + C_2^{(1)}(eB)\langle O_{2}\rangle&=&
   \frac{1}{\pi}\int_0^{s_0}
   \!\!\!\!\!ds\, s
   {\mbox{Im}}\,\Pi_0^{\mbox{\tiny{HAD}}} (s)\nn
   &-&
   \frac{C_0}{2}s_0^2 + C_0^{(2)}(eB)^2.
\label{two}
\end{eqnarray}

In order to  set up explicitly the relevant FESR we start by computing the hadronic contribution. The axial-vector current in the presence of a magnetic field can be interpolated by the charged pion current
\bea
  A_\mu=-f_\pi D_\mu \pi^+ 
 = -f_\pi (\partial_\mu - ie \cal{A}_\mu)\pi^+,
 \label{interpol}
\eea
where $f_\pi = 130.28(14)$ MeV~\cite{PDG} is the pion decay constant, $\pi^+$ the pion field, and  ${\cal{A}}_\mu=(B/2)(0,-y,x,0)$  the vector potential in the symmetric gauge, which gives rise to a constant magnetic field along the $\hat{z}$ direction. Therefore, the axial-vector correlator in the hadronic sector can be written as
\bea
   \Pi^{\mathrm{\mbox{\tiny{HAD}}}}_{\mu\nu}(x,y) &\equiv& <0|T( A_{\mu}(x) \;, \; A_{\nu}^{\dagger}(y))|0> \nn
    &=& i f_\pi^2 \langle 0|T\left[D_\mu\pi^+(x)D^*_\nu\pi^-(y)\right]|0\rangle \nn
 &=& if_\pi^2D_\mu(x)D_\nu^*(y)G_\pi(x,y)\nn
 &=& ie^{ie\Phi(x,y)}f_\pi^2\left(\partial/\partial x^\mu\right)\left(\partial/\partial y^\mu\right)
 \widetilde{G}_\pi(x-y)\nn
 &\equiv& e^{ie\Phi(x,y)}\widetilde{\Pi}^{\mathrm{\mbox{\tiny{HAD}}}}_{\mu\nu}(x-y),
\label{hadsect}
\eea
where we have used the fact that the charged pion propagator $G_\pi(x,y)$ in the presence of a magnetic field can be written as a product of a transnationally invariant piece $\widetilde{G}_\pi(x-y)$ and a {\it phase} factor $e^{ie\Phi(x,y)}$. The phase factor does not depend on the integration path so that choosing a straight line trajectory it can be written as
\bea
 \Phi(x,y)=\int_x^y {\cal{A}}(\xi) d\xi,
 \label{phasefac}
\eea
where $\xi=yt+x(1-t)$, $t\in [0,1]$. It is easy to show that the above phase factor can be {\it gauged away} by a suitable gauge transformation of the vector potential. Hence, we keep only the translational invariant part of the hadronic correlator whose Fourier transform is
\bea
 \Pi^{\mathrm{\mbox{\tiny{HAD}}}}_{0}(q^2)= i  f_\pi^2 \widetilde{G}_\pi(q^2),
\label{Fourier}
\eea
where $\widetilde{G}_\pi(q^2)$ stands for the Fourier transform of $\widetilde{G}_\pi(x-y)$.
Using Schwinger's proper time method this quantity can be written as
\bea
\widetilde{G}_\pi(q^2) = \int_0^\infty \frac{d\tau}{\cos(eB\tau)}
 e^{i\tau[q_\parallel^2-q_\perp^2 \tan (eB\tau)/eB \tau +i\epsilon]},
 \label{Schwinger}
\eea
where  $m_\pi=0$ as we consider the chiral limit. Hereafter we shall use the notation 
\bea
g_{\mu\nu}&=&g_{\mu\nu}^\parallel - g_{\mu\nu}^\perp\nn
g^\parallel_{\mu\nu} &=& \textrm{diag}(1,0,0,-1)\nn
g^\perp_{\mu\nu} &=& \textrm{diag}(0,1,1,0).
\label{gs}
\eea 
Consequently
\bea
a\cdot b &=& (a\cdot b)_\parallel - (a\cdot b)_\perp\nn
(a\cdot b)_\parallel &=& g^\parallel_{\mu\nu}a^\mu b^\nu\nn
&=& a_0b_0-a_3b_3\nn
(a\cdot b)_\perp &=&g^\perp_{\mu\nu}a^\mu b^\nu\nn
&=& a_1b_1+a_2b_2\nn
g^\parallel_{\mu\nu}g^{\mu\nu} &=& 2\nn
g^\perp_{\mu\nu}g^{\mu\nu} &=& -2.
\label{prods}
\eea
Extending the integration in Eq.~(\ref{Schwinger}) to the lower complex plane, it can be shown~\cite{Mex} that the charged pion propagator is expressed in terms of a sum over Landau levels
\bea
 \widetilde{G}_\pi(q^2) = 2i\sum_{l=0}^\infty
 \frac{(-1)^l L_l(2q_\perp^2/eB)e^{-q_\perp^2/eB}}{q_\parallel^2-(2l+1)eB}.
 \label{sum}
\eea

As it is customary in problems where Lorentz invariance is lost and the correlators depend separately on the time and space parts of the four-momentum, to simplify the analysis hereafter we study the behavior of the correlator in the {\it static} limit where the space part of the four momentum vanishes. However, note that since the magnetic field separates space-time into longitudinal and transverse directions with respect to the direction of the field, $q_0$ comes together with $q_3$ to form $q^2_\parallel= q_0^2 - q_3^2$. It is therefore equally simple to set $q_{\perp}^2=0$, perform the analysis in terms of $q^2_\parallel$ and then at the end, if needed, take $q_3\rightarrow0$. In this limit,  Eq.~(\ref{sum}) becomes
\bea
    \widetilde{G}_\pi(q^2) = 2i\sum_{l=0}^\infty
    \frac{(-1)^l}{q_\parallel^2-(2l+1)eB}.
\label{afterqT0}
\eea
Therefore the hadronic contribution to the correlator becomes explicitly
\bea
 \Pi^{\mathrm{\mbox{\tiny{HAD}}}}_{0}(q_\parallel^2=s)= -2  f_\pi^2 \sum_{l=0}^\infty
 \frac{(-1)^l}{s-(2l+1)eB}.
\label{afterqT0expl}
\eea
As we show in Appendix A, the imaginary part of Eq.~(\ref{afterqT0expl}) in the weak field limit, $eB < s_0$, is given by
\bea
   {\mbox{Im}}\Pi^{\mathrm{\mbox{\tiny{HAD}}}}_{0}(s)=f_\pi^2\pi\delta(s-eB).
\label{imofafterqT0expl}
\eea
It is important to mention that Eq.~(\ref{imofafterqT0expl}) holds provided $s_0$ satisfies also the condition $s_0 < 3eB$, as  explicitly shown in Appendix A. This means that the weak field condition ($eB<s_0$) has to be supplemented with a further restriction for Eq.~(\ref{imofafterqT0expl}) to remain valid. Recall that $s_0$ represents the onset for the pQCD description for the axial-vector spectral density and that this quantity is a decreasing function of temperature~\cite{di-muon}. For cold nuclear matter, as in the case of a neutron star, the condition $eB < s_0 < 3eB$ may be difficult to meet, specially for a weak field where not only the situation $eB<s_0$, but even $neB<s_0$, ($n\geq 1$) can happen. However, for a heavy-ion collision, around the deconfinement/chiral symmetry restoration transition, when $s_0$ has dropped off to small values, the weak field condition can also be made compatible with $eB<s_0<3eB$. Hereafter we keep in mind this last observation as the working scenario, aiming to eventually incorporate thermal effects to describe the behavior of a magnetized medium near the phase transition. 

Using Eq.~(\ref{imofafterqT0expl}), the hadronic line integral in the FESR is given explicitly by
\bea
   \frac{1}{\pi}\int_0^{s_0}
   \!\!\!\!\!ds\, s^{N-1}   
   {\mbox{Im}}\Pi_0^{\mbox{\tiny{HAD}}} (s)= f_\pi^2\,\left( eB \right)^{N-1}.
\label{hadcontfin}
\eea
Substituting Eq.~(\ref{hadcontfin}) into the QCD sum rules Eqs.~(\ref{one})-(\ref{two}) gives
\bea
   0&=&f_\pi^2 - C_0s_0 + C_0^{(1)}(eB)\nn
   -C_4\langle O_4\rangle&=&f_\pi^2 (eB) -  \frac{C_0}{2}s_0^2 + C_0^{(2)}(eB)^2.
\label{firsttwosumrules}
\eea
In order to solve these equations, we now proceed to compute explicitly the coefficients $C_0^{(1)}$ and $C_0^{(2)}$.

\begin{figure}
{\centering {
\includegraphics[scale=0.5]{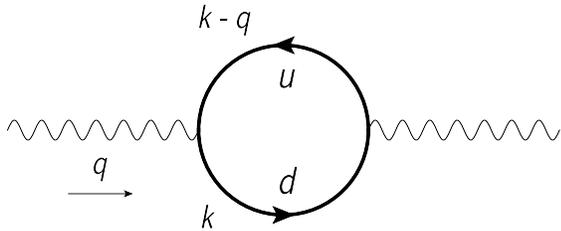}
}}
\caption{pQCD contribution to the axial-vector current correlator in the presence of a magnetic field. The thick internal lines represent the full quark propagators in the magnetic field background.}
\label{fig2}
\end{figure}

\section{Solutions for $s_0$ and $C_4\langle O_4\rangle$}\label{III}

To perform the perturbative calculation of the coefficients $C_0^{(1)}$ and $C_0^{(2)}$ in the weak field limit we make use of the weak field expansion of the quark propagator in the presence of a constant magnetic field~\cite{Chyi}, and  in the chiral limit, up to order ${\mathcal{O}}(B^2)$
\bea
   iS_B(k)&=&i\frac{\slsh{k}}{k^2} - (e_qB) \frac{\gamma_1\gamma_2(\gamma\cdot k)_\parallel}{k^4}\nn
   &-&2i(e_qB)^2\frac{\left[k_\perp^2(\gamma\cdot k)_\parallel - 
   k_\parallel^2(\gamma\cdot k)_\perp \right]}{k^8},
\label{propB}
\eea
where $e_q$ is the absolute value of the quark's charge. We emphasize that the weak field expansion of the the fermion propagator in the chiral limit is a well defined object. This happens because the weak field limit of the propagator can be thought of as a series representation in powers of $eB$ of the full propagator independent of any relation between the strength of the magnetic field and the fermion mass. One can then use this series representation and take the chiral limit as much as one can do it for the usual fermion propagator in the absence of the field; depending on the process considered, this limit could lead to infrared divergences but it does not preclude the validity of taking $m\rightarrow0$ inside the propagator. In the present context, the field can be considered as weak when compared to $s_0$, which is the only other energy (squared) present in the problem, in such a way that the expansion in Eq.~(\ref{pQCD}) converges.

\begin{figure}
{\centering {
\includegraphics[scale=0.5]{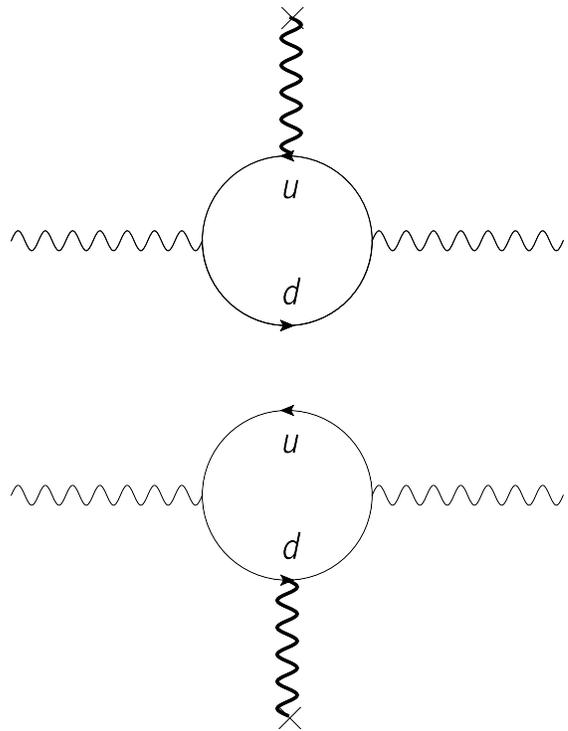}
}}
\caption{pQCD contribution to the axial-vector current correlator in the presence of a magnetic field to first order in $e_qB$. The thick wavy lines ending in a cross represent the external magnetic field.}
\label{fig3}
\end{figure}
The pQCD contribution to the axial-vector current correlator in the presence of a magnetic field is depicted in Fig.~\ref{fig2}, where we also define the kinematics. The thick internal lines represent the full quark propagators in the magnetic field background. To first order in $e_qB$ only one of the two quark propagators carries the magnetic effects. This is depicted in Fig.~\ref{fig3} where the wavy line starting from a cross represents the external magnetic field. The two diagrams in Fig.~\ref{fig3} that determine the coefficient $C_0^{(1)}$, vanish identically when contracted with the momenta carried by the axial-vector currents. This is due to a straightforward application of Furry's theorem, and to the fact that  the vector and axial-vector correlators are chiral symmetric. However, we  show explicitly in Appendix B that each of the contributions in the diagrams, Fig.~\ref{fig3}, vanishes.

\begin{figure}
{\centering {
\includegraphics[scale=0.5]{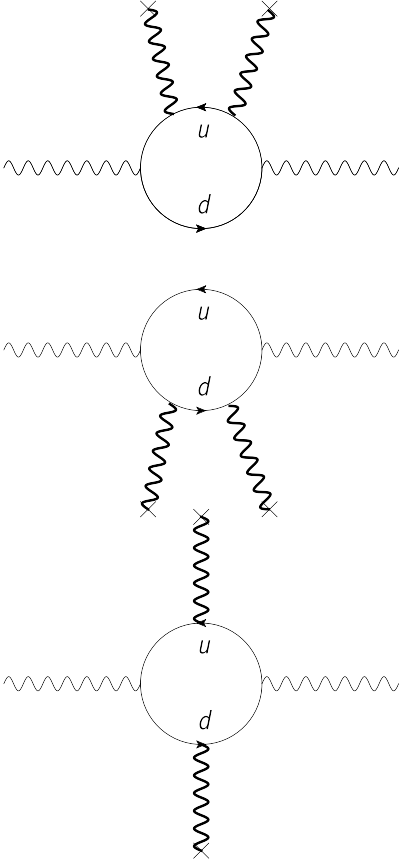}
}}
\caption{pQCD contribution to the axial-vector current correlator in the presence of a magnetic field to second order in $e_qB$. The thick wavy lines ending in a cross represent the external magnetic field.}
\label{fig4}
\end{figure}

The first non-trivial magnetic contribution to the pQCD axial-vector current correlator is of order $(e_qB)^2$. The relevant diagrams are shown in Fig.~\ref{fig4}. First, we compute the diagram where one magnetic field line is attached to each one of the two quark propagators. For these, we use Eq.~(\ref{propB}) to first order in $e_qB$. We call this contribution $\Pi_{\mu\nu}^{(11)}(q^2)$, and its explicit expression is
\bea
   &&\Pi_{\mu\nu}^{(11)}(q^2) =- i N_c(q_uq_dB^2)\nn
   &\times&
   \int\frac{d^4k}{(2\pi)^4}\frac{{\mbox{Tr}}\left[\gamma_\mu\gamma_1\gamma_2
   \left[\gamma\cdot (k - q)_\parallel\right]
   \gamma_\nu\gamma_1\gamma_2(\gamma\cdot k)_\parallel\right]}{(k-q)^4k^4}.\nn
\label{Pi11}
\eea

Since according to Eq.~(\ref{Cauchy}), we are interested in the magnetic corrections to the coefficient of the longitudinal structure, $\Pi_0(q^2)$ we project $\Pi_{\mu\nu}^{(11)}(q^2)$ with $q^\mu q^\nu$ and define
\bea
  \widetilde{\Pi}_{0}^{(11)}(q^2) \equiv q^{\mu}q^\nu\Pi_{\mu\nu}^{(11)}(q^2).
\label{magmodPi11}
\eea
Using 
\bea
   \gamma_1\gamma_2
   \left[\gamma\cdot (k - q)_\parallel\right]=\left[\gamma\cdot (k - q)_\parallel\right]
   \gamma_1\gamma_2
\label{firstidentity}
\eea
together with
\bea
   \gamma_1\gamma_2 \slsh q\gamma_1\gamma_2=2(\gamma\cdot q)_\perp - \slsh q,
\label{secondidentity}
\eea
gives
\bea
   \widetilde{\Pi}_{0}^{(11)}&\!\!\!\!=\!\!\!\!&iN_c(q_uq_dB^2)\int\frac{d^4k}{(2\pi)^4}\nn
   &\!\!\!\!\times\!\!\!\!&
   \frac{{\mbox{Tr}}\left[
   \slsh q\left[\gamma\cdot (k-q)\right]_\parallel(-2(\gamma\cdot q)_\perp + \slsh q)(\gamma\cdot k)_\parallel\right]
   }{(k-q)^4k^4}.\nn
\label{magmod-2Pi11}
\eea
The evaluation of the trace yields
\bea
   &&{\mbox{Tr}}\left[
   \slsh q\left[\gamma\cdot (k-q)\right]_\parallel(-2(\gamma\cdot q)_\perp + \slsh q)(\gamma\cdot k)_\parallel\right]\nn
   &&
   =-4(q_\parallel^2 + q_\perp^2)\left[(k-q)\cdot k\right]_\parallel\nn
   &&+\ 8(q\cdot k)_\parallel\left[(k-q)\cdot q\right]_\parallel.
\label{trace}
\eea
We now use the Feynman parametrization 
\bea
   \frac{1}{(k-q)^4k^4}=3!\int_0^1dx\frac{x(x-1)}{\left[(k-xq)^2-x(x-1)q^2\right]^4},
\label{Feynman}
\eea
and the change of variable
\bea
   k\rightarrow l=k-xq,
\label{change}
\eea
to obtain
\bea
   \widetilde{\Pi}_{0}^{(11)}(q^2)&\!\!\!=\!\!\!&4iN_c(q_uq_dB^2)3!\int_0^1x(x-1)\int\frac{d^4l}{(2\pi)^4}
   \frac{1}{\left[l^2-\Delta\right]^4}\nn
   &\!\!\!\times\!\!\!&
   \left[
   q^2q_\parallel^2x(x-1) - (q_\parallel^2 + q_\perp^2)l_\parallel^2 + 2(q\cdot l)_\parallel^2
   \right],
\label{withl}
\eea
where we have discarded terms with odd powers of $\sl{l}$ and  defined $\Delta=x(x-1)q^2$. The integrals over $\sl{l}$ are computed by means of
\bea
   \int\frac{d^dl}{(2\pi)^d}\frac{1}{\left[l^2-\Delta\right]^n}&=&i\frac{(-1)^n}{(4\pi)^{d/2}}
   \frac{\Gamma (n-d/2)}{\Gamma (n)}\left(\frac{1}{\Delta}\right)^{n-d/2}\nn
   \int\frac{d^dl}{(2\pi)^d}\frac{l^\mu l^\nu}{\left[l^2-\Delta\right]^n}&=&i\frac{(-1)^{n-1}}{(4\pi)^{d/2}}
   \frac{g^{\mu\nu}}{2}
   \frac{\Gamma (n-d/2-1)}{\Gamma (n)}\nn
   &\times&\left(\frac{1}{\Delta}\right)^{n-d/2-1},
\label{integrals}
\eea
with $n=4$ and $d=4$. Using Eq.~(\ref{integrals}) in Eq.~(\ref{withl}), and after integrating over $x$, we find
\bea
   \widetilde{\Pi}_{0}^{(11)}&\!\!\!=\!\!\!&-\frac{N_c}{4\pi^2}(q_uq_dB^2)\frac{\left[q_\parallel^2 + q_\perp^2\right]}{q^2}.
\label{almostPi011}
\eea
In the limit $q_\perp^2\rightarrow 0$ Eq.~(\ref{almostPi011}) becomes
\bea
   \widetilde{\Pi}_{0}^{(11)}&\stackrel{q_\perp^2\rightarrow 0}{\longrightarrow}&-\frac{N_c}{4\pi^2}(q_uq_dB^2).
\label{finPi011}
\eea
In a similar fashion we compute the diagrams in Fig.~\ref{fig4} to second order in $eB$ in the $u$-quark and in the $d$-quark propagator. Calling the longitudinal projections $\widetilde{\Pi}_0^{(20)}$ and $\widetilde{\Pi}_0^{(02)}$, respectively, the result is
\bea
   \widetilde{\Pi}_0^{(20)}(q^2)&=&-\frac{N_c}{24\pi^2}(q_uB)^2\left[\frac{(q_\parallel^2 + q_\perp^2)}{q^2}
   +2\frac{q_\parallel^2q_\perp^2}{q^4}\right]\nn
   &\stackrel{q_\perp^2\rightarrow 0}{\longrightarrow}&-\frac{N_c}{24\pi^2}(q_uB)^2\nn
   \widetilde{\Pi}_0^{(02)}(q^2)&=&-\frac{N_c}{24\pi^2}(q_dB)^2\left[\frac{(q_\parallel^2 + q_\perp^2)}{q^2}
   +2\frac{q_\parallel^2q_\perp^2}{q^4}\right]\nn
   &\stackrel{q_\perp^2\rightarrow 0}{\longrightarrow}&-\frac{N_c}{24\pi^2}(q_dB)^2.
\label{theothers}
\eea 
Adding all three contributions, and using the absolute values  $q_u=2/3\ e$, $q_d=1/3\ e$,  and $N_c=3$, we obtain  the coefficient of the longitudinal structure of the axial-vector current correlator to second order in the magnetic field
\bea
   \Pi_0^{B^2}=-\left(\frac{17}{18}\right)\frac{(eB)^2}{4\pi^2}.
\label{dongbfield}
\eea
Using this result together with  the first equation in  Eq.~(\ref{pQCD}), we obtain the Wilson coefficient of the pQCD contribution to second order in the magnetic field
\bea
    C_0^{(2)}=-\left(\frac{17}{18}\right)\frac{1}{4\pi^2}.
\label{coefC02}
\eea
\begin{figure}
{\centering {
\includegraphics[scale=0.5]{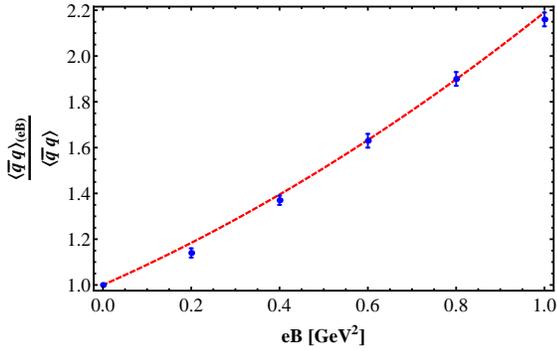}
}}
\caption{(Color on-line) Light-quark condensate normalized to its vacuum value as a function of the magnetic field strength $eB$ in units of GeV$^2$. The data points are from Ref.~\cite{Bali} and the dotted line corresponds to the fit $\langle \bar{q}q\rangle_{(eB)}/\langle \bar{q}q\rangle = 1 + a (eB) + b (eB)^2$, with $a=0.85$ GeV$^{-2}$, $b=0.34$ GeV$^{-4}$.}
\label{fig5}
\end{figure}
The last ingredient needed to find $s_0$ and $C_4\langle O_4\rangle$  is the magnetic field dependence of $f_\pi$. This may be obtained from the Gell-Mann-Oakes-Renner (GMOR) relation, which relates $f_\pi$ with the light-quark condensate, and the pion mass with the light-quark masses. 
\begin{figure}[b!]
{\centering {
\includegraphics[scale=0.5]{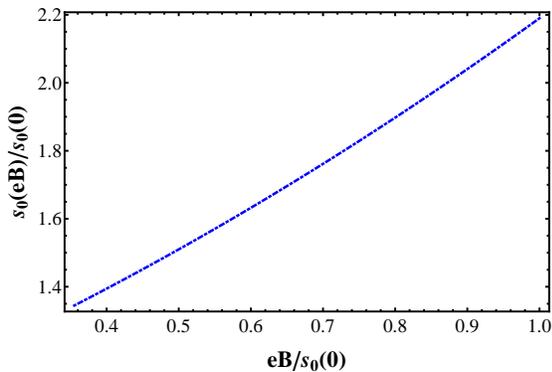}
}}
\caption{(Color on-line) Squared energy pQCD threshold, $s_0$, normalized to its $eB=0$ value, as a function of the field strength $eB$.  Note that the plotted range is consistent with the requirement $eB < s_0 < 3eB.$}
\label{fig6}
\end{figure}
The GMOR relation is a low energy theorem involving the pseudoscalar current correlator
\begin{equation}
\psi_5(q^2) = i \int\, d^{4} x \, e^{i q x} 
 <0|T( \partial^\mu A_{\mu}(x) \, \partial^\nu A_{\nu}^{\dagger}(0))|0>, 
\end{equation}
where the divergence of the axial-vector current, $\partial^\mu A_{\mu}(x)$, is proportional to the sum of the up- and down-quark masses. At zero momentum the GMOR relation reads
\begin{equation}
m_\pi^2\,f_\pi^2= - 2\,(m_u + m_d) \langle \bar{q} q \rangle, 
 \label{explicitGMOR}
\end{equation}
In the standard chiral interpretation of this relation $f_\pi$ is related to the light quark condensate 
\bea
f_\pi^2 = - 2{\mathcal{B}}\langle \bar{q} q \rangle,
\label{GMOR}
\eea
and the pion mass is related to the light-quark masses
\begin{equation}
m_\pi^2 = \frac{1}{\mathcal{B}}(m_u + m_d),
 \label{massesGMOR}
\end{equation}
where ${\mathcal{B}}$ is a CHPT constant. In order to find the constant $\cal{B}$  one  needs to use  the physical pion and light-quark masses. In the hadronic sector the pion pole contribution to the pseudoscalar current correlator, $\psi_5(q^2)$ is given by
\begin{equation}
\psi_5^{\mbox{\tiny{HAD}}}(q^2)= if_{\pi}^2m_{\pi}^4\widetilde{G}_\pi(q^2),
 \label{hadgmor}
\end{equation}
where $\widetilde{G}_\pi(q^2)$ is the pion propagator. In the QCD sector the light-quark condensate contribution to $\psi_5(0)$ is given by
\begin{equation}
\psi_5^{\mbox{\tiny{QCD}}}(0)=-2(m_u+m_d)\langle \bar{q}q \rangle.
 \label{qcdgmor}
\end{equation}
According to Eq.~(\ref{afterqT0}) [see also the discussion leading to Eq.~(\ref{delta}) for the normalization] the pion propagator for $q^2\geq 0$ in the presence of a weak magnetic field can be approximated as
\begin{equation}
\widetilde{G}_\pi(q^2) \simeq \frac{i}{q^2-m_\pi^2-eB},
\label{accordingref}
\end{equation}
so that the GMOR relation  becomes
\begin{equation}
\frac{f_\pi^2 m_\pi^4}{m_\pi^2+eB}\simeq -2(m_u + m_d)\langle\bar{q} q\rangle.
\label{gmorbecomes}
\end{equation}
For weak fields, we use the expansion
\begin{equation}
\frac{1}{m_\pi^2+eB} \sim \frac{1}{m_\pi^2} (1 - \frac{eB}{m_\pi^2} ).
\label{weakgmor}
\end{equation}
Inserting Eq.~(\ref{weakgmor}) into Eq.~(\ref{gmorbecomes}), and neglecting a magnetic field dependence of the pion and quark masses,  the term $\frac{eB}{m_\pi^2}$ introduces a correction to the GMOR relation of higher order than the one being considered in this work. Therefore, to a good approximation for small fields  the magnetic field dependence of $f_\pi^2$ is determined by that of $<\bar{q} q>$ and the constant ${\mathcal{B}}$ from its vacuum value. We point out that the validity of the GMOR relation in the presence of a magnetic field has first been shown in the first of Refs.~\cite{SHUSH}. 

\begin{figure}
{\centering {
\includegraphics[scale=0.5]{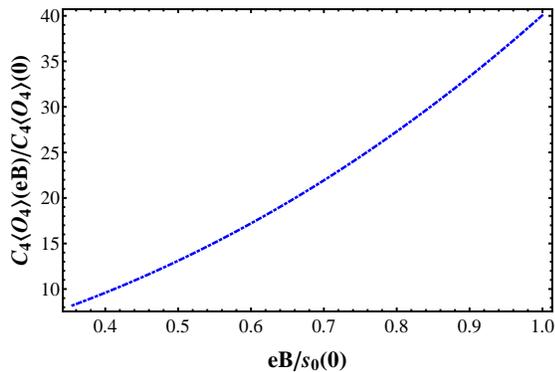}
}}
\caption{(Color on-line) The gluon condensate, $C_4\langle O_4\rangle$, normalized to its $eB=0$ value, as a function of $eB$. Note that the plotted range is consistent with the requirement $eB < s_0 < 3eB.$}
\label{fig7}
\end{figure}

The light-quark condensate in the presence of the magnetic field has been computed in Ref.~\cite{Bali}. We make use of this result, and parametrize the magnetic field dependence of the light-quark condensate with a quadratic fit
\bea
   \langle \bar{q}q\rangle_{(eB)}/\langle \bar{q}q\rangle = 1 + a (eB) + b (eB)^2,
\label{fit}
\eea
where $a=0.85$ GeV$^{-2}$, $b=0.34$ GeV$^{-4}$ and $(eB)$ is given in GeV$^2$.
The data from Ref.~\cite{Bali} together with the fit are shown in Fig.~\ref{fig5}. Using this information we finally write the explicit solutions for  $s_0$ and $C_4\langle O_4\rangle$ from Eq.~(\ref{firsttwosumrules})
\bea
   s_0&=&-8\pi {\mathcal{B}}\langle \bar{q}q\rangle_{(eB)}\nn
   C_4\langle O_4\rangle&=& -2(eB) {\mathcal{B}}\langle \bar{q}q\rangle_{(eB)}
   +  8 \pi ({\mathcal{B}}\langle \bar{q}q\rangle_{(eB)})^2\nn
   &+&\left(\frac{17}{18}\right)\frac{(eB)^2}{4\pi^2}.
\label{firsttwosumrulesexpl}
\eea
The solutions for $s_0$ and for $C_4\langle O_4\rangle$ as functions of $eB$ are plotted in Figs.~\ref{fig6} and~\ref{fig7}, respectively.  Note that $s_0$ is proportional to the absolute value of the light-quark condensate, and that together with $C_4\langle O_4\rangle$ it increases with increasing magnetic field.

\section{Summary and conclusions}\label{concl}

In this paper we studied QCD FESR for the axial-vector current correlator in the presence of a magnetic field in the weak field limit $eB< s_0$, where $s_0$ is the squared energy threshold for the onset of pQCD. Given that we work in the massless quark limit, the weak field limit can be understood as an expansion in the small parameter $eB/s_0$. We have shown that the presence of the field modifies both the pQCD as well as the hadronic sectors of the FESR. The direct modification to the hadronic contribution for the equation that governs the behavior of $s_0$ vanishes non-trivially. The only change in this  sector comes from the dependence of the pion decay constant on the magnetic field. In turn, this pion decay constant is proportional to the quark condensate through the  GMOR relation. We have shown that in the limit where the magnetic field is small compared to the physical pion mass squared, the proportionality constant between $f_\pi$ and $\langle\bar{q}q\rangle$ can be approximated to be the same as in vacuum. Use of lattice QCD data on the quark condensate as a function of the magnetic field allows then to extract the magnetic field dependence of $f_\pi$. We should point out that since lattice QCD results are valid for physical pion masses, the calculation in this work cannot be directly compared to the results in Refs.~\cite{SHUSH} which are performed using chiral perturbation theory in the limit where $qB\gg m_\pi^2$. The magnetic field dependence of $s_0$ is proportional to the magnetic field dependence of the absolute value of the light-quark condensate. This means that the behavior of these two parameters as a function of the magnetic field is similar. Therefore the magnetic field both helps the formation of the condensate and acts against deconfinement. Next, we computed explicitly  the magnetic corrections  to the pQCD contribution which affect the behavior of the gluon condensate as a function of field strength. In the weak field limit the first correction is quadratic in the field.  The gluon condensate also grows as a function of the field strength which goes hand in hand with the behavior of the magnetic field, both as a catalyst of chiral symmetry breaking and confinement. 

The results obtained here should serve as a basis for studies at finite temperature in an external magnetic field, thus providing  clues on the relation between chiral symmetry restoration and deconfinement at the transition temperature. This  research is  in progress and will be reported elsewhere.

\section*{Acknowledgments}

The authors acknowledge useful conversations with A. Raya during the genesis of this project. This work was supported in part by the National Research Foundation (South Africa), the Oppenheimer Memorial Trust OMT Ref. 20242/02, UNAM-DGAPA-PAPIIT grant number 101515,  CONACyT-Mexico grant number 128534, FONDECYT (Chile) grant Nos. 130056 and 1120770, and 1150847.

\section*{Appendix A: Imaginary part of the hadronic contribution to the axial-vector current correlator}

The charged axial-vector current correlator in the presence of a constant magnetic field, and in the limit 
$q_\perp \rightarrow 0$ is written as
\bea
 \Pi_0^{\mbox{\tiny{HAD}}}(s)=-2f_{\pi}^2 \sum_{l=0}^{\infty}\frac{(-1)^l}{s-(2l+1)eB}.
\eea
The above expression can be split into two sums, one for the even and the other for the odd values of $l$, namely
\bea
 \Pi_0^{\mbox{\tiny{HAD}}}(s)&=&-2f_{\pi}^2 \left\{ \sum_{l=0,even} \frac{1}{s-(2l+1)eB}\right.
\nonumber \\
 &-&\left. \sum_{l=odd}\frac{1}{s-(2l+1)eB} \right\}.
 \label{corr+-}
\eea
Pulling out a factor $-1/4eB$ from both sums and adding and substracting the element with
$l=-1$ we obtain from Eq.~(\ref{corr+-})
\bea
 \Pi_0^{\mbox{\tiny{HAD}}}(s)&=&2f_{\pi}^2 \left\{\frac{1}{4eB}
\sum_{l'=0}^{\infty}\frac{1}{l'-\frac{s/eB-1}{4}}\right. \nonumber \\
 &-& \left. \frac{1}{4eB}\sum_{l'=0}^{\infty}
\frac{1}{l'-\frac{s/eB+1}{4}}-\frac{1}{s+eB}\right\},
 \label{corrlprime}
\eea
where we defined $l'=\frac{l}{2}$ for the sum with even $l$ and $l'=\frac{l+1}{2}$ for the sum with odd $l$. 
The sums in Eq.~(\ref{corrlprime}) are divergent. In order to extract the finite piece we regularize them as
\bea
  \Pi_0^{\mbox{\tiny{HAD}}}(s)&=&2f_{\pi}^2 
  \lim_{\epsilon\rightarrow 1}
  \left\{\frac{1}{4eB}
\sum_{l'=0}^{\infty}\frac{1}{(l'-\frac{s/eB-1}{4})^{\epsilon}}\right. \nonumber \\
 &-& \left. \frac{1}{4eB}\sum_{l'=0}^{\infty}
\frac{1}{(l'-\frac{s/eB+1}{4})^{\epsilon}}-\frac{1}{s+eB}\right\}\nn
&=&2f_{\pi}^2 
  \lim_{\epsilon\rightarrow 1}
  \left\{\frac{1}{4eB}
\zeta(\epsilon,(s/eB - 1)/4)\right.\nn
 &-&\left. \frac{1}{4eB}\zeta(\epsilon,(s/eB + 1)/4)-\frac{1}{s+eB}\right\}
, \nonumber \\
 \label{correregul}
\eea
where $\zeta(a,z)$ is the Hurwitz zeta function. Expanding around $\epsilon=1$ we find
\bea
 \Pi_0^{\mbox{\tiny{HAD}}}(s)&=&2f_{\pi}^2 \left\{ \frac{1}{4eB}\left[
 \frac{1}{\epsilon - 1} -
 \psi\left( \frac{-(s/eB-1)}{4}
\right)\right] \right. \nonumber \\
 &-& \frac{1}{4eB}\left[
 \frac{1}{\epsilon - 1} -
 \psi\left( \frac{-(s/eB+1)}{4} \right)\right]\nn
 &-&\left. \frac{1}{s+eB}
\right\}\nn
&=&-2f_{\pi}^2 \left\{ \frac{1}{4eB}\psi\left( \frac{-(s/eB-1)}{4}
\right) \right. \nonumber \\
 &+& \left.\frac{1}{4eB}
 \psi\left( \frac{-(s/eB+1)}{4} \right) + \frac{1}{s+eB}
\right\},
 \label{finalcorre}
\eea
where $\psi(x)$ is the di-gamma function. We note that the divergent pieces cancel when $\epsilon\rightarrow 1$. Recall that $\psi(x)$ is singular for $x=0,-1,-2,\ldots$. In the region $0\leq eB < s_0$, neither of the di-gamma functions in Eq.~(\ref{finalcorre}) becomes singular. The first singularity for $\psi(-(s/eB-1)/4)$ happens at $s=eB$ and for $\psi(-(s/eB+1)/4)$ at $s=3B$. Therefore, by restricting the analysis to 
the region $eB\leq s_0<3eB$ we can compute the discontinuity, or imaginary part of Eq.~(\ref{finalcorre}), with the result
\bea
 {\mbox{Im}}\Pi_0^{\mbox{\tiny{HAD}}}(s)=f_{\pi}^2 \pi \delta(s-eB),
 \label{imaginarycorre}
\eea
where since $s$ is strictly larger or equal to 0, one has
\bea
   \lim_{\epsilon\rightarrow 0}\frac{\epsilon}{(s-eB)+\epsilon^2}=\frac{\pi}{2}\delta(s-eB).
\label{delta}
\eea
Finally, in the limit $eB \rightarrow 0$ the imaginary part of the correlator becomes
\bea
 {\mbox{Im}}\Pi_0^{\mbox{\tiny{HAD}}}(s)=f_{\pi}^2 \pi \delta(s),
 \label{imaginarycorrezero}
\eea
which coincides with the known value in the absence of a magnetic field.

\section*{Appendix B: First order magnetic correction to the pQCD contribution to the axial-vector current correlator}

The contribution to the pQCD axial-vector current correlator of order $(eB)$ is given by
\bea
   \Pi_{\mu\nu}^{B}(q^2)= \Pi_{\mu\nu}^{(10)}(q^2)+\Pi_{\mu\nu}^{(01)}(q^2)
\label{PiB}
\eea
where
\bea
   \Pi_{\mu\nu}^{(10)}(q^2)&=& N_c (q_uB)\int \frac{d^4k}{(2\pi)^4}\frac{{\mbox{Tr}}\left[\gamma_1\gamma_2
   [\gamma\cdot (k-q)_\parallel]\gamma_{\mu}\slsh k \gamma_{\nu} \right]}{k^2(k-q)^4}\nn
   \Pi_{\mu\nu}^{(01)}(q^2)&=& N_c (q_dB)\int \frac{d^4k}{(2\pi)^4}\frac{{\mbox{Tr}}\left[(\slsh k - \slsh q)\gamma_{\mu}\gamma_1\gamma_2[\gamma\cdot q]_\parallel \gamma_{\nu} \right]}{(k-q)^2k^4}\nn
\label{Pi0110}
\eea
Note that the traces in Eqs.~(\ref{Pi0110}) are equal except for an overall sign,  due to the ordering of elements inside the trace. Therefore, we  evaluate one of the traces, 
\bea
   {\mbox{Tr}}\left[\gamma_1\gamma_2[\gamma\cdot (k-q)_\parallel]\gamma_{\mu}\slsh k \gamma_{\nu} \right]
   &=&k^{\alpha}(k-q)^{\beta}_\parallel\nn
   &=& {\mbox{Tr}}\left[ \gamma_\mu \gamma_\alpha \gamma_\nu \gamma_1 \gamma_2 \gamma_\beta^\parallel \right]\nn
   &=&-ik^{\alpha}{\mbox{Tr}}\left[ \gamma_\mu \gamma_\alpha \gamma_\nu \gamma_\rho\gamma_5\right.\nn
   &\times&\left\{b^\rho[(k-q)\cdot u] \right.\nn
   &-&\left. \left. u^\rho[(k-q)\cdot b]\right\}\right],
\label{traceb}
\eea
where $u^\rho$ and $b^\rho$ are four-vectors describing the particle's rest frame and the direction of the magnetic field, respectively. In the rest frame, there are given by
\bea
   u&=&(1,0,0,0)\nn
   b&=&(0,0,0,1).
\label{vectors}   
\eea
Using that
\bea
   {\mbox{Tr}}\left[ \gamma_\mu \gamma_\alpha \gamma_\nu \gamma_\rho \gamma_5 \right]=-4i \epsilon_{\mu \alpha \nu \rho},
\label{theorem} 
\eea
we can write Eq.~(\ref{traceb}) as
\bea
   &&{\mbox{Tr}}\left[\gamma_1\gamma_2[\gamma\cdot (k-q)_\parallel]\gamma_{\mu}\slsh k \gamma_{\nu} \right]=\nn
   &&
   -4k^\alpha \epsilon_{\mu \alpha \nu \rho} \left\{ b^\rho [(k-q)\cdot u]-u^\rho [(k-q)\cdot b] \right\}.
\label{trace10}
\eea
Notice that when Eq.~(\ref{trace10}) is contracted with $q^\mu q^\nu$ the coefficient of the longitudinal structure vanishes.

\end{document}